\documentclass[referee]{raa}
\usepackage{graphicx,times}
\usepackage{natbib}
\usepackage{amssymb,amsmath}

\usepackage{enumitem}

\defcitealias{bp82}{BP82}
\defcitealias{c02}{C02}
\defcitealias{fbg04}{FBG04}
\defcitealias{fgr10}{FGR10}
\defcitealias{gwl09}{GWL09}
\defcitealias{mr06}{MR06}
\defcitealias{nm12}{NM12}

\begin{document}

   \title{A disk-corona model for low/hard state of black hole X-ray binaries
$^*$
\footnotetext{\small $*$ Supported by the National Natural Science Foundation of China.}
}

 \volnopage{ {\bf 2013} Vol.\ {\bf X} No. {\bf XX}, 000--000}
   \setcounter{page}{1}

   \author{Jiu-Zhou Wang\inst{1}, Ding-Xiong Wang\inst{1} and Chang-Yin Huang\inst{1}
   }

   \institute{School of Physics, Huazhong University of Science and Technology, 430074, Wuhan, China; {\it dxwang@mail.hust.edu.cn}\\
\vs \no
   {\small Received ; accepted }
}

\abstract{A disk-corona model for fitting the low/hard (LH) state of the associated steady jet in black hole
X-ray binaries (BHXBs) is proposed based on the large-scale magnetic field configuration that arises from the
coexistence of the Blandford-Znajek (BZ) and Blandford-Payne (BP) processes, where the
magnetic field configuration for the BP process is determined by the requirement of energy
conversion from Poynting energy flux into kinetic energy flux in the jet. It is found that corona
current is crucial to guarantee the consistency of the jet launching from the accretion disk. The
relative importance of the BZ and BP processes in powering jets from black hole accretion disks is
discussed, and the LH state of several BHXBs is fitted based on our model. In addition, we
suggest that magnetic field configuration can be regarded as the second parameter for governing
the state transition of BHXBs.
}

   \authorrunning{J.-Z. Wang et al. }            
   \titlerunning{A disk-corona model for low/hard state}  
   \maketitle

%
\section{Introduction}           
\label{sect1}

Spectral states observed in black hole X-ray binaries (BHXBs) involve a number of
unresolved issues in astrophysics and display complex variations not only in the luminosities and
energy spectra, but also in the presence/absence of jets and quasi-periodic oscillations (QPOs). Not
long ago, \citet[hereafter \citetalias{mr06}]{mr06} used four parameters to define X-ray
states based on the very extensive RXTE data archive for BHXBs, in which three states, i.e.,
thermal–dominant state, low/hard (LH) state and steep power law state are included. Although a
consensus on classification of spectral states of BHXBs has not been reached, it is widely accepted
that these states can be reduced to only two basic states, i.e., a hard state and a soft one, and jets
can be observed in hard states, but cannot be in soft states.

The accretion flow in LH state is usually supposed to be a truncated thin disk with an inner
advection-dominated accretion flow (ADAF) in the prevailing scenario (\citealt{emn97,eea98};
\citetalias{mr06}; \citealt{dgk07}). Generally speaking, the thermal component of the
spectra of BHXBs can be well fitted by a truncated thin accretion disk, while the power law
component can be interpreted by an ADAF. Although the X-ray, EUV, and UV spectra of XTE
J1118+480 can be satisfactorily explained by a truncated thin disk plus an ADAF (\citealt{eea01}),
the IR fluxes are significantly underestimated and the radio emission cannot be
interpreted. \citet{ycn05} fitted the spectrum of XTE J1118+480, and proposed a coupled
accretion-jet model to interpret the observations, in which the jet dominates the radio and infrared
emission, the thin disk dominates the UV emission, and the hot flow produces most of the X-ray
emission. This model successfully fits the multiwavelength spectrum of the source, and further testing
of this model can be seen in \citet{zyc10}.

An ADAF plus a truncated thin disk has become the major model used in interpreting spectra of
BHXBs in LH state; however, recent observations show some contradiction with it. For example,
\textit{XMM-Newton} observations of GX 339-4 show that a broad iron line together with a dim, hot
thermal component was present in its spectra during the hard state. This effect seems to be
observed in a few other sources such as Cygnus X-1 and SWIFT J1753.5-0127 (\citealt{mhm06,mea06}).
Recently, \citet{rmf09} studied the Chandra observation of XTE J1118+480 in the canonical LH state,
and a thermal disk emission with a temperature of approximately 0.21keV is found at greater than the $14\sigma$ confidence level, and they concluded that this thermal emission most likely originates from an accretion disk extending close to innermost stable circular orbit (ISCO). The results of fits made to both components (thermal component and broad iron line) strongly suggest that a standard thin disk remains at or near to ISCO, at least in bright phases of
LH state.

In order to interpret the thermal component and broad iron line in the luminous LH state,
some authors suggested that the accretion geometry could be described as a cool inner disk and an
even cooler outer disk, separated by a gap filled with an ADAF (\citealt{mp07}; \citealt{lea07}).

Recently, \citet{rfm10} presented an X-ray study of eight black holes (BHs) in LH state, and they found that a thermal disk continuum with a color temperature consistent with $L\propto T^4$ is clearly detected in all eight sources and the detailed fits to the line profiles exclude a truncated disk in each case.

Besides the power-law component dominates, another feature of LH state of BHXBs is its
association with quasi-steady jets. Although ADAF model is successful in fitting the spectra of LH
state of some BHXBs, the detail of how associated jets are produced has not been addressed.

Different mechanisms have been proposed to interpret the jet production in BH systems
of different scales, such as the plasma gun (\citealt{c95}), the cosmic battery (\citealt{ck98}) and
the magnetic tower (\citealt{l96}), the most promising mechanisms for powering jets are Blandford-Znajek (BZ) and Blandford-Payne (BP) processes, which relies on a poloidal, large-scale magnetic field anchored on an accretion disk around a spinning BH (\citealt[hereafter \citetalias{bp82}]{bz77,bp82}; \citealt{l02,dea12}; for a review see \citealt{s10}).

In this paper, we intend to model LH state of BHXBs based on a disk-corona model, in which
the inner edge of the accretion disk is assumed to extend to ISCO, and the jets are driven by the
large-scale open magnetic field of the coexistence of the BZ and BP processes. This paper is
organized as follows. In section 2, based on the energy conversion from Poynting energy flux into
the kinetic energy flux in the jet, we argue that some current within corona is required to flow
across the magnetic surfaces, which are formed due to the rotation of the open field lines anchored
at the accretion disk. Henceforth the current is referred to as corona current. In section 3, we
propose the magnetic field configuration of the coexistence of the BZ and BP processes based on
the energy conversion in the jet, and discuss the relative importance of these two mechanisms in
driving jets from BH systems. In section 4, the spectral profiles of the LH state of four BHXBs are
fitted based on our model, and the relation between jet power and X-ray luminosity is checked by
adjusting accretion rate and the outer boundary of the BP magnetic field configuration. Finally, in
section 5, we discuss some issues related to our model. We propose a scenario of state transitions
from LH state to very high (VH) state, and suggest that the magnetic field configuration could be
regarded as the second parameter in state transitions of BHXBs.

Throughout this paper the geometric units $G=c=1$ are used.

\section{CONVERSION OF ENERGY IN JETS AND CORONA CURRENT}
\label{sect2}

\begin{figure}[ht]
\centering
\includegraphics[width=120mm]{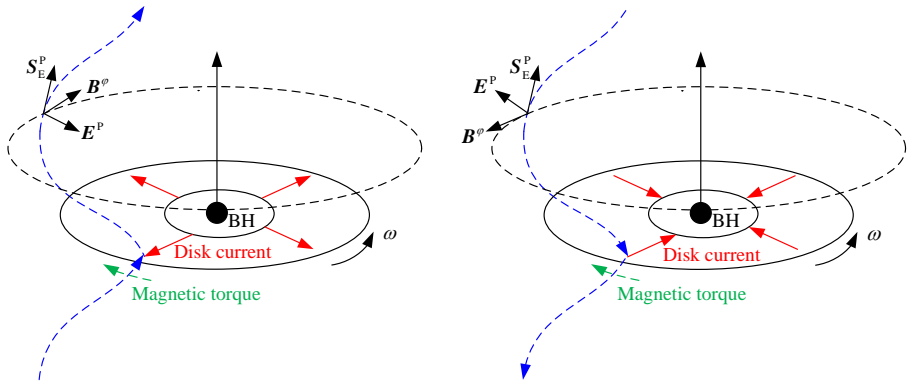}
\caption{Poynting flux is driven by a rotating disk around a BH via magnetic torque exerted on disk current.
    The green arrow represents the direction of magnetic torque, and the red solid and the blue dashed arrows
    represent disk current and magnetic field lines, respectively.}
\label{fig1}
\end{figure}

Both matter outflow and Poynting flux are produced via the large-scale magnetic field
anchored on the disk around a rotating BH. What is the relation between the two kinds of the
fluxes? As shown in Figure \ref{fig1}, Poynting flux $\mathbf{S}_\mathrm{E}^\mathrm{P}
=\mathbf{E}^\mathrm{P}\times\mathbf{B}^\varphi$ is produced due to the magnetic
field lines dragged by the rotating disk, where $\mathbf{E}^\mathrm{P}$ is the poloidal
induced electric field, and $\mathbf{B}^\varphi$ is the toroidal magnetic field.
Obviously, both $\mathbf{E}^\mathrm{P}$ and $\mathbf{B}^\varphi$ arise from disk rotation,
and they are expressed as follows,
\begin{eqnarray}
\mathbf{E}^\mathrm{P} & = & -\mathbf{v}^\mathrm{F}\times\mathbf{B}^\mathrm{P},\label{eq1}\\
\mathbf{S}_\mathrm{E}^\mathrm{P} & = & \mathbf{E}^\mathrm{P}\times\mathbf{B}^\varphi,\label{eq2}
\end{eqnarray}
where $\mathbf{S}_\mathrm{E}^\mathrm{P}$ is the poloidal Poynting flux along the field line.

According to \citetalias{bp82}, the conservation of energy and angular momentum along each field line
can be written as follows,
\begin{eqnarray}
e & = & e_\mathrm{matter}+e_\mathrm{Poynting}=const,\label{eq3}\\
l & = & l_\mathrm{matter}+l_\mathrm{Poynting}=const.\label{eq4}
\end{eqnarray}

The quantities $e_\mathrm{matter}$ and $e_\mathrm{Poynting}$ are specific energies of matter
and electromagnetic (EM) field, respectively, and they read (\citetalias{bp82})
\begin{equation}
\label{eq5}
\left\{ \begin{array}{ll}
e_\mathrm{matter}=v^2/2+h+\Phi,\\
e_\mathrm{Poynting}=-\omega r B^\varphi/k,
\end{array} \right.
\end{equation}
where $r$ is the cylindrical radius of the field line, and $\omega$ is the angular velocity of the field line,
which is equal to the angular velocity of the disk $\Omega_\mathrm{d}=\frac{1}{M(\chi^3+a_*)}$
at the radius of the footpoint $r_\mathrm{d}=M\chi^2$. The quantities $l_\mathrm{matter}$ and $l_\mathrm{Poynting}$
are respectively specific angular momenta of matter and EM field, and they read
\begin{equation}
\label{eq6}
\left\{ \begin{array}{ll}
l_\mathrm{matter}=r v^\varphi,\\
l_\mathrm{Poynting}=-r B^\varphi/k,
\end{array} \right.
\end{equation}
where the quantities $h$, $\Phi$ and $-\omega r B^\varphi/k$ in equation (\ref{eq5}) are specific enthalpy,
gravitational potential, and the work done on the streaming gas by the magnetic torque, respectively. The
quantity $-r B^\varphi/k$ in equation (\ref{eq6}) represents the impulse of the magnetic torque, and the
parameter $k$ is related to the ratio of the mass flux to the magnetic flux for each magnetic field line
as follows,
\begin{equation}
\label{eq7}
k/4\pi \equiv \rho v^\mathrm{P}/B^\mathrm{P}.
\end{equation}

The meanings of $e_\mathrm{Poynting}$ and $l_\mathrm{Poynting}$ can be clarified more clearly as follows.
The poloidal flux of EM angular momentum can be written as
$S_\mathrm{L}^\mathrm{P}=-r B^\varphi B^\mathrm{P}/4\pi=-r B^\varphi\rho v^\mathrm{P}/k$
(\citealt{mt82}), thus we have
\begin{equation}
\label{eq8}
\left\{ \begin{array}{ll}
\frac{S_\mathrm{L}^\mathrm{P}}{\rho v^\mathrm{P}}=-r B^\varphi/k=l_\mathrm{Poynting},\\
\frac{S_\mathrm{E}^\mathrm{P}}{\rho v^\mathrm{P}}=-\omega r B^\varphi/k=e_\mathrm{Poynting}.
\end{array} \right.
\end{equation}

We conclude that $e_\mathrm{Poynting}$ and $l_\mathrm{Poynting}$ are respectively
EM specific energy and angular momentum corresponding to mass flux. Based on Ampere’s law we have
\begin{equation}
\label{eq9}
\oint \mathbf{B}\cdot d\mathbf{l}=2\pi r B^\varphi=4\pi\sum I.
\end{equation}

As shown in equation (\ref{eq8}), $e_\mathrm{Poynting}$ is proportional to $rB^\varphi$. Considering that
$e_\mathrm{Poynting}$ is converted to kinetic energy continuously in the jet (\citetalias{bp82}; \citealt{s96,s10}), we infer that the absolute values of both $rB^\varphi$ and $\sum I$ in equation (\ref{eq9}) must decrease continuously along the jet, where $\sum I$ is the algebraic sum of current flowing inside the magnetic surface formed due to the rotation of the field line.

\begin{figure}[ht]
\centering
\includegraphics[width=85mm]{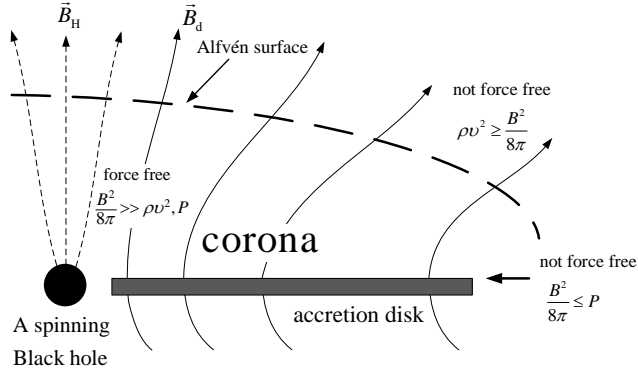}
\caption{Three regions in a magnetically accelerated flow from an accretion disk. Corona is assumed to exist
between the disk surface and Alfven surface indicated by the dashed thick line. $B_\mathrm{H}$ (dashed thin lines) and $B_\mathrm{d}$ (solid thin lines) represent the poloidal magnetic field on the BH horizon and disk, respectively.}
\label{fig2}
\end{figure}

In standard model for jet launched by magneto-centrifugal acceleration there are three
distinct regions as shown in Figure \ref{fig2} (\citealt{br76}, \citetalias{bp82}; \citealt{s96,s10}).

\begin{figure}[ht]
\centering
\includegraphics[width=120mm]{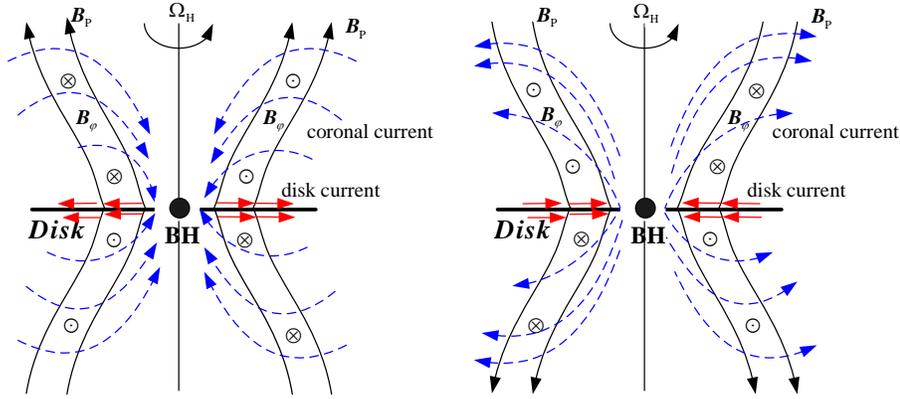}
\caption{A schematic drawing of corona current flowing across magnetic surface, where disk and corona
    currents are represented respectively by red solid and blue dashed arrows. The symbols $\odot$
    and $\otimes$ represent the outward and inward toroidal magnetic field lines, respectively.}
\label{fig3}
\end{figure}

In the atmosphere of the disk up to the Alfven surface the magnetic field dominates over gas
pressure and kinetic energy of the outflow, and the outflow experiences a centrifugal force accelerating
along the field lines in this region of force free. On the other hand, corona is a
perfect launching site for outflow from accretion disk (\citealt{mf02}), and disk-corona
model provides a possible scenario for interpreting LH state associated with a quasi-steady jet
from BHXBs. From the above discussion, we infer that corona current must flow across the
magnetic surfaces as shown in Figure 3, and it can be expressed from equation (\ref{eq9}) as follows,
\begin{equation}
\label{eq10}
I_\mathrm{cor}(r)=r B^\varphi/2,
\end{equation}
where $I_\mathrm{cor}(r)$ is the corona current threading the magnetic surface above the cylindrical radius $r$.
Inspecting Figure \ref{fig3}, we find that corona current is essential to interpret energy conversion in
the jet.

There are two puzzles related to corona current. The first one is whether the corona current
can flow across the magnetic surface in the region of centrifugal acceleration, where $B^2/8\pi\gg\rho v^2$
is required as shown in Figure \ref{fig2}. In fact, the condition for centrifugal acceleration does not imply no current flowing across the magnetic surface. Inspecting Figures \ref{fig1} and \ref{fig3}, we find that the corona current is required by the continuity of the current flowing in the disk, and it is driven by the induced electric field $\mathbf{E}^\mathrm{P}$ or the electric potential difference between the two adjacent magnetic surfaces.

The second puzzle is that the quantity $-\omega r B^\varphi/k$ appears to have two different meanings,
i.e., (i) the work done on the streaming gas by the magnetic torque \citepalias{bp82}, and (ii) EM specific
energy $e_\mathrm{Poynting}$ along a field line given by equation (\ref{eq8}). How to understand the work done by the magnetic torque decreases continuously during the energy conversion in the jet? This puzzle can
be resolved easily by invoking corona current. The work done by the magnetic torque consists of
two parts, one is on the disk current, and the other one is on the corona current. From Figure \ref{fig3} we find that the two works done by the magnetic torque have opposite signs, because the direction of the
disk current is opposite to that of the corona current. The total work by the magnetic torque is the
integral of the differential work from the neutral plane at $z = 0$ to the Alfven surface. So the
work is zero at $z = 0$ for $B^\varphi =0$, and it attains its maximum at the disk surface, and then it
decreases along the jet due to the negative work on the corona current. It is the work done on the corona current by the magnetic torque that gives rise to the conversion of EM energy into kinetic energy in the jet.

Thus we conclude that corona current is not only required by the continuity of the disk
current but also is essential for understanding energy conversion in the jet. In addition, we can estimate the efficiency of the conversion from EM energy into kinetic energy in the jet in terms of $r B^\varphi$. The conversion efficiency can be defined as the ratio of $e_\mathrm{matter,A}$ to $e$, which are the specific energy of matter at the Alfven surface and the total specific energy along a field line, respectively. Thus we have conversion efficiency as
\begin{eqnarray}
 \eta_\mathrm{E} & \equiv & e_\mathrm{matter,A}/e=(e-e_\mathrm{Poynting,A})/e \nonumber \\
   & \simeq & 1-e_\mathrm{Poynting,A}/e_\mathrm{Poynting,d}=1-(r B^\varphi)_\mathrm{A}/(r B^\varphi)_\mathrm{d},\label{eq11}
\end{eqnarray}
where $e_\mathrm{Poynting,d}$ and $e_\mathrm{Poynting,A}$ are the EM specific energy at disk surface and Alfven surface, respectively. In deriving the above equation, $e_\mathrm{Poynting,d}\simeq e$ is assumed, since EM specific energy is dominant at disk surface.

Thus, we infer that the conversion efficiency depends on the variation of $r B^\varphi$ along the
field line. For example, we have about $1/3$ EM energy converted into the kinetic energy in the jet
for the ratio $(r B^\varphi)_\mathrm{A}/(r B^\varphi)_\mathrm{d}=2/3$.

\section{MAGNETIC FIELD CONFIGURATION BASED ON ENERGY CONVERSION}
\label{sect3}

We can constrain the magnetic field configuration in accretion disk based on the energy
conversion in the jet. The power of the magnetic torque on the radial disk current between the two
adjacent magnetic surfaces is
\begin{equation}
\label{eq12}
d\mathrm{P}_\mathrm{d}=B_\mathrm{d}^\mathrm{P}\Omega_\mathrm{d} I_\mathrm{d} r_\mathrm{d} dr_\mathrm{d},
\end{equation}
where the subscript `d' indicates the quantities at disk surface. On the other hand, the work done
on the streaming gas per unit mass at the cylindrical radius $r$ is
\begin{equation}
\label{eq13}
W_\mathrm{line}(r)=-\omega r B^\varphi/k=2\omega I_\mathrm{cor}(r)/k,
\end{equation}
Incorporating equations (\ref{eq10}) with (\ref{eq13}), and considering $\omega=\Omega_\mathrm{d}$, we have
\begin{equation}
\label{eq14}
d\mathrm{P}_\mathrm{d}=W_\mathrm{line}(r_\mathrm{d})\dot{M}_\mathrm{jet}dr_\mathrm{d}
=(2\omega I_\mathrm{cor}(r_\mathrm{d})/k)\dot{M}_\mathrm{jet}dr_\mathrm{d},
\end{equation}
\textbf{where $\dot{M}_\mathrm{jet}$ is the mass outflow rate in the jet of unit width, being expressed in equation (\ref{eq17}).}

Considering the continuity of the corona current and disk current, we have
$I_\mathrm{cor}(r_\mathrm{d})=I_\mathrm{d}(r_\mathrm{d})$. Incorporating equations
(\ref{eq12}) and (\ref{eq14}), we have the relation between mass loss rate at $r_\mathrm{d}$ and the
poloidal magnetic field $B_\mathrm{d}^\mathrm{P}$ as follows,
\begin{equation}
\label{eq15}
\dot{M}_\mathrm{jet}(r_\mathrm{d})=B_\mathrm{d}^\mathrm{P} r_\mathrm{d}k/2.
\end{equation}

Following \citet{bb99}, we have accretion rate $\dot{M}$ varying with the disk radius as follows,
\begin{equation}
\label{eq16}
\dot{M}=\dot{M}_\mathrm{in}(r_\mathrm{d}/r_\mathrm{in})^s, \;\;\;\;\;\;\; 0<s<1,
\end{equation}
where $\dot{M}_\mathrm{in}$ is the accretion rate at the inner edge of the disk, which is related to
Eddington luminosity by $\dot{M}_\mathrm{in}=\dot{m}_\mathrm{in}L_\mathrm{Edd}/(0.1c^2)$. henceforth the subscript `in'
indicates the quantities at the inner edge of the accretion disk. The mass outflow rate in the jet is given by
\begin{equation}
\label{eq17}
\dot{M}_\mathrm{jet}(r_\mathrm{d})=d\dot{M}/d r_\mathrm{d}
=\dot{M}_\mathrm{in}(s/r_\mathrm{in})(r_\mathrm{d}/r_\mathrm{in})^{s-1},
\end{equation}

Incorporating equations (\ref{eq15}) and (\ref{eq17}), we have the relation between poloidal magnetic field at disk surface and $\dot{M}_\mathrm{in}$ as follows,
\begin{equation}
\label{eq18}
B_\mathrm{d}^\mathrm{P}(r_\mathrm{d})=\dot{M}_\mathrm{in}(2s/k r_\mathrm{in}^2)(r_\mathrm{d}/r_\mathrm{in})^{s-2}.
\end{equation}

The poloidal magnetic field far from the disk surface is assumed to be roughly self-similar, and is
given as (\citetalias{bp82}, \citealt{lpp94}),
\begin{equation}
\label{eq19}
B^\mathrm{P}(r_\mathrm{d},\varsigma)=B_\mathrm{d}^\mathrm{P}(r_\mathrm{d})\varsigma^{-\alpha},
\end{equation}
where $\varsigma\equiv (r/r_\mathrm{d})$ is the cylindrical radius of the field line. Incorporating equations
(\ref{eq18}) and (\ref{eq19}), we have  the 3-D axisymmetric magnetic field distribution on the accretion disk as follows,
\begin{equation}
\label{eq20}
B_\mathrm{d}^\mathrm{P}(r_\mathrm{d},\varsigma)=B_\mathrm{in}(r_\mathrm{d}/r_\mathrm{in})^{s-2}\varsigma^{-\alpha},
\end{equation}
where $B_\mathrm{in}$ is the poloidal magnetic field at the inner edge of the disk.

The strength of the magnetic field on the BH horizon can be determined based on the balance
between the magnetic pressure on the horizon and the ram pressure of the innermost parts of an
accretion as follow (\citealt{msl97}),
\begin{equation}
\label{eq21}
B_\mathrm{H}^2/(8\pi)=P_\mathrm{ram} \sim \rho c^2 \sim \dot{M}_\mathrm{in}/(4\pi r_\mathrm{H}^2),
\end{equation}
Equation (\ref{eq21}) can be rewritten as
\begin{equation}
\label{eq22}
\dot{M}_\mathrm{in}=\alpha_\mathrm{m} B_\mathrm{H}^2 r_\mathrm{H}^2=\alpha_\mathrm{m} (1+q)^2 B_\mathrm{H}^2 M^2,
\end{equation}
where $r_\mathrm{H} \equiv M(1+q)$ is the radius of BH horizon, and $q \equiv \sqrt{1-a_*^2}$ is a function of BH spin,
and the parameter $\alpha_\mathrm{m}$ is adjustable due to the uncertainty of equation (\ref{eq22}).

The optimal BZ power is given by equation (\ref{eq23}) as a function of BH spin (\citealt{lwb00,wxl02}),
and the BP power is given by equation (\ref{eq24}) as an integral over the region with large-scale open magnetic field from the inner edge to the outer boundary (\citealt[hereafter \citetalias{c02}]{c02}).
\begin{equation}
\label{eq23}
\left\{ \begin{array}{ll}
\mathrm{P}_\mathrm{BZ}=B_\mathrm{H}^2 M^2 Q^{-1}(\arctan Q-a_*/2)  \\
Q\equiv \sqrt{(1-q)/(1+q)}
\end{array} \right.,
\end{equation}
\begin{equation}
\label{eq24}
\mathrm{P}_\mathrm{BP}=\int_{r_\mathrm{in}}^{r_\mathrm{out}}(\gamma_\mathrm{j}-1)\dot{M}_\mathrm{jet}d r_\mathrm{d}
=\dot{M}_\mathrm{in}s \int_1^{\xi_\mathrm{out}}(\gamma_\mathrm{j}-1)\xi^{s-1}d\xi ,
\end{equation}
where $\xi_\mathrm{out} \equiv r_\mathrm{out}/r_\mathrm{in}$ is the radius of the outer boundary
of the large scale open magnetic field in terms of $r_\mathrm{in}$. The parameter $\gamma_\mathrm{j}\equiv (1-v_\mathrm{A}^2)^{-1/2}$
is the Lorentz factor of the outflow at Alfven surface, and it is related to the parameters $\alpha_\mathrm{m}, s, a_*$ and $\alpha$ by
\begin{equation}
\label{eq25}
\frac{\xi^{s-2}\chi_\mathrm{in}^4}{\alpha_\mathrm{m}s(1+q)^2}\left(\frac{\xi \chi_\mathrm{in}^2}{\xi^{3/2}\chi_\mathrm{in}^3+a_*}\right)^\alpha
=\gamma_\mathrm{j}^{-\alpha}(\gamma_\mathrm{j}^2-1)^{(\alpha+1)/2},
\end{equation}
where $\chi_\mathrm{in}$ is defined as $\chi_\mathrm{in} \equiv \sqrt{r_\mathrm{in}/M}$. The derivation of equation (\ref{eq25})
is given in Appendix.

The relative importance of the BZ and BP processes can be estimated by incorporating equations
(\ref{eq23}), (\ref{eq24}) with (\ref{eq22}) based on the magnetic field configuration given in Figure \ref{fig1}, and the
ratio of the BZ to BP powers is
\begin{equation}
\label{eq26}
\mathrm{P}_\mathrm{BZ}/\mathrm{P}_\mathrm{BP}
=\frac{Q^{-1}(\arctan Q-a_*/2)}{\alpha_\mathrm{m}(1+q)^2 s \int_1^{\xi_\mathrm{out}}(\gamma_\mathrm{j}-1)\xi^{s-1}d\xi}.
\end{equation}

\begin{figure}[ht]
\includegraphics[width=144mm]{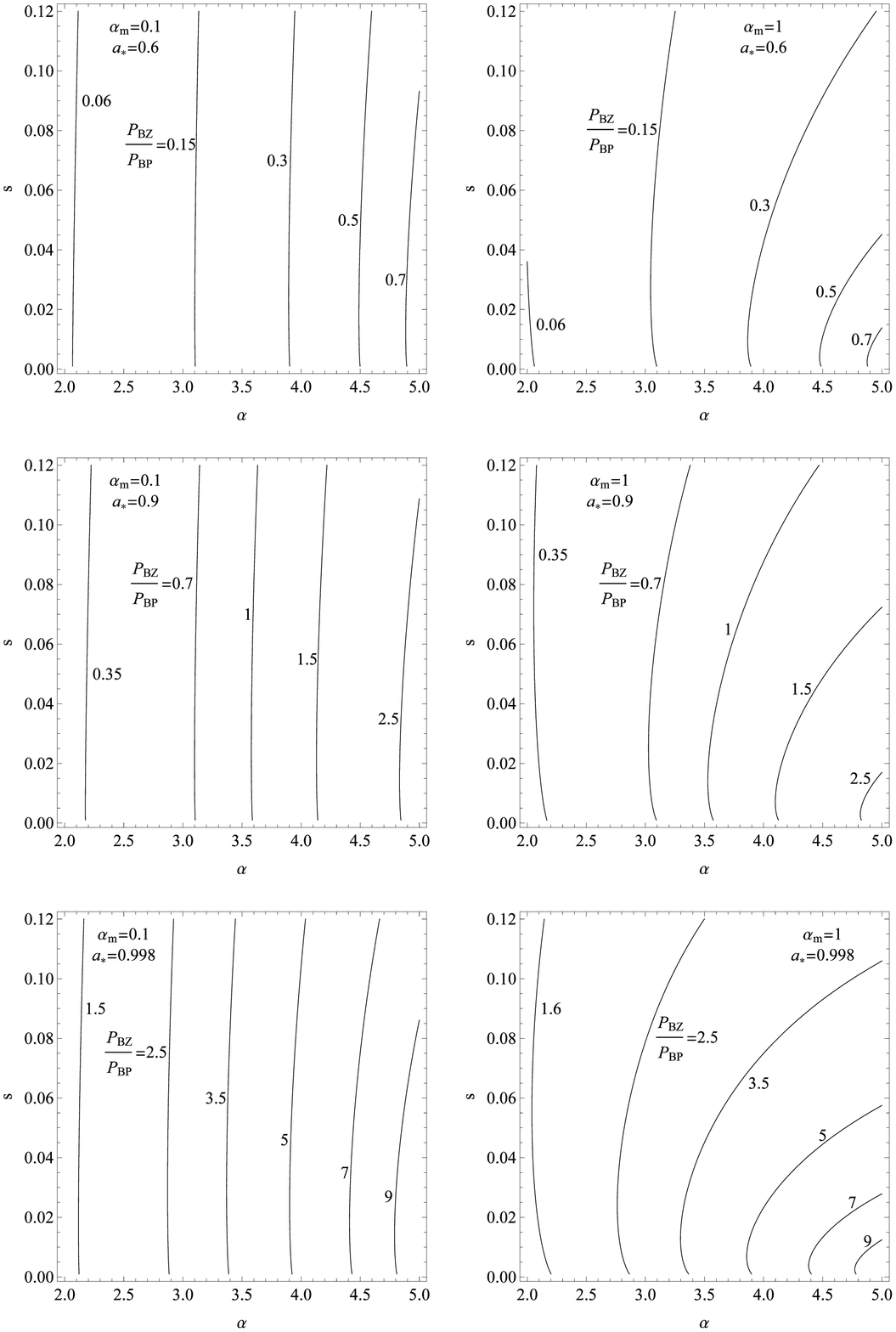}
\caption{The contours of the ratio of $P_\mathrm{BZ}$ to $P_\mathrm{BP}$ in $\alpha$ -- s parameter space with different values of $\alpha_\mathrm{m}$ and $a_*$.}
\label{fig4}
\end{figure}

Four parameters ($\alpha_\mathrm{m}$, $a_*$, $s$ and $\alpha$) are involved in equation (\ref{eq26}), and
$r_\mathrm{out}=1000M$ is fixed in calculations. By using equation (\ref{eq26}) we have the contours
of the ratio of $\mathrm{P}_\mathrm{BZ}$ to $\mathrm{P}_\mathrm{BP}$ in $\alpha - s$ parameter space
with different values of $\alpha_\mathrm{m}$ and $a_*$ as shown in Figure \ref{fig4}.

Inspecting Figure \ref{fig4}, we find that the ratio of $\mathrm{P}_\mathrm{BZ}$ to $\mathrm{P}_\mathrm{BP}$
is less than or around unity for $0 < s < 0.12$, and $2 < \alpha \leq 5$ with $\alpha_\mathrm{m}=0.1,1$. It implies that the BZ power is not dominant over the BP power for the great outer boundary of the open magnetic field on the disk, $r_\mathrm{out}=1000M$, except the extreme BH spin $a_* \rightarrow 0.998$ with $\alpha \sim 5$, and this result is in accordance with those obtained by other authors (e.g., \citealt{ga97,lop99,m99}).

\section{FITTINGH LH STATE OF FOUR BHXBS}
\label{sect4}

In this section we intend to fit the LH state associated with quasi-steady jets of four BHXBs,
XTE J1550−564, GRO J1655−40, GRS 1915+105 and 4U 1543−47, and the jet power is regarded as the sum of the BZ and BP powers, i.e.,
\begin{equation}
\label{eq27}
\mathrm{P}_\mathrm{jet}=\mathrm{P}_\mathrm{BZ}+\mathrm{P}_\mathrm{BP}.
\end{equation}

In addition, we discuss the constraints of the relation between jet power and X-ray luminosity on the variation of $\dot{m}_\mathrm{in}$ and $r_\mathrm{out}$ in state transition of BHXBs.

\subsection{Effect of Jet Launching from Accretion Disk on Energy and Angular Momentum}
\label{sect4.1}

The fitting of LH state is given based on the conservation of energy and angular momentum
by considering jet launching from the accretion disk. Following \citetalias{c02}, the kinetic flux of the jet can
be written as
\begin{equation}
\label{eq28}
F_\mathrm{jet}=\dot{m}_\mathrm{jet}(\gamma_\mathrm{j}-1).
\end{equation}

Considering that Poynting flux is much larger than kinetic flux near disk surface, we can
relate $F_\mathrm{jet}$ at Alfven surface to the Poynting flux at the disk surface as follows,
\begin{equation}
\label{eq29}
S_\mathrm{E}^\mathrm{P}=3F_\mathrm{jet},
\end{equation}
where the factor `3' in equation (\ref{eq29}) implies that one third energy in Poynting flux is assumed to
be converted into kinetic energy of the jet.

As is well known, the angular momentum flux $S_\mathrm{L}^\mathrm{P}$ extracted electromagnetically from the
disk surface is related to the Poynting energy flux as follows,
\begin{equation}
\label{eq30}
S_\mathrm{L}^\mathrm{P}=S_\mathrm{E}^\mathrm{P}/\Omega_\mathrm{d},
\end{equation}
Incorporating equations (\ref{eq28})--(\ref{eq30}), we have
\begin{equation}
\label{eq31}
S_\mathrm{L}^\mathrm{P}=3\dot{m}_\mathrm{jet}(\gamma_\mathrm{j}-1)/\Omega_\mathrm{d},
\end{equation}
where $\dot{m}_\mathrm{jet} \equiv \dot{M}_\mathrm{jet}/4\pi r_\mathrm{d}$
is the mass loss rate per unit area at the footpoint of the jet.

The integrated shear stress of the disk should be affected unavoidably by the transport of
angular momentum and energy in the jet, resulting in the decrease of the disk dissipation and disk
radiation. At the presence of the jet the conservation equations of energy and angular momentum
can be written as
\begin{eqnarray}
\frac{d}{d r_\mathrm{d}}(\dot{M}_\mathrm{d}E^\dagger-T_\mathrm{visc}\Omega_\mathrm{d})
& = & 4\pi r_\mathrm{d}[(\dot{m}_\mathrm{jet}+F_\mathrm{rad})E^\dagger+S_\mathrm{L}^\mathrm{P}\Omega_\mathrm{d}],\label{eq32}\\
\frac{d}{d r_\mathrm{d}}(\dot{M}_\mathrm{d}L^\dagger-T_\mathrm{visc})
& = & 4\pi r_\mathrm{d}[(\dot{m}_\mathrm{jet}+F_\mathrm{rad})L^\dagger+S_\mathrm{L}^\mathrm{P}],\label{eq33}
\end{eqnarray}
where $T_\mathrm{visc}$ and $F_\mathrm{rad}$ are respectively the internal viscous torque and the energy flux radiated
away from the surface of disk, $E^\dagger$ and $L^\dagger$ are respectively specific energy and angular
momentum of the disk matter, being expressed by (\citealt{nt73})
\begin{eqnarray}
E^\dagger & = & (1-2\chi^{-2}+a_* \chi^{-3})/(1-3\chi^{-2}+2a_* \chi^{-3})^{1/2},\label{eq34}\\
L^\dagger & = & M\chi(1-2a_*\chi^{-3}+a_*^2 \chi^{-4})/(1-3\chi^{-2}+2a_* \chi^{-3})^{1/2},\label{eq35}
\end{eqnarray}
where $\chi \equiv \sqrt{r_\mathrm{d}/M}=\xi^{1/2} \chi_\mathrm{in}$, and the quantities $E^\dagger$ and $L^\dagger$ are related by
\begin{equation}
\label{eq36}
d E^\dagger/d r_\mathrm{d}=\Omega_\mathrm{d} d L^\dagger/d r_\mathrm{d}.
\end{equation}

Incorporating equations (\ref{eq32}), (\ref{eq33}) and (\ref{eq36}), we have the radiation flux from disk as follows,
\begin{eqnarray}
F_\mathrm{rad}(r_\mathrm{d}) & = & -\frac{d\Omega_\mathrm{d}/d r_\mathrm{d}}{4\pi r_\mathrm{d}}
(E^\dagger-\Omega_\mathrm{d} L^\dagger)^{-2}
\times \Big(\int_{r_\mathrm{in}}^{r_\mathrm{d}}(E^\dagger-\Omega_\mathrm{d} L^\dagger)\dot{M}\frac{d L^\dagger}
{d r_\mathrm{d}}d r_\mathrm{d} \nonumber \\
& & +\,(E^\dagger_\mathrm{in}-\Omega_\mathrm{d,in} L^\dagger_\mathrm{in})T_\mathrm{in}
-\int_{r_\mathrm{in}}^{r_\mathrm{d}}(E^\dagger-\Omega_\mathrm{d} L^\dagger)4\pi r_\mathrm{d} S_\mathrm{L}^\mathrm{P}d r_\mathrm{d}
\Big),\label{eq37}
\end{eqnarray}
where $E^\dagger_\mathrm{in}$, $L^\dagger_\mathrm{in}$, $\Omega_\mathrm{d,in}$ and $T_\mathrm{in}$ in equation (\ref{eq37}) are respectively specific energy, specific angular momentum, angular velocity and torque at the inner edge of the accretion disk.

Inspecting equation (\ref{eq37}), we find that jet launched from accretion disk does result in a
negative contribution on the disk radiation, which is represented by the term related to the angular
momentum flux $S_\mathrm{L}^\mathrm{P}$. Thus we think that jet launching from accretion disk is indeed essential for interpreting the associated of LH state with quasi-steady jet in BHXBs.

Furthermore, we obtain a rather tight constraint on the parameters $s$, $\alpha$, $\alpha_\mathrm{m}$ and $\dot{m}_\mathrm{in}$, involved in our model based on the following arguments.

\begin{figure}[ht]
\centering
\includegraphics[width=144mm]{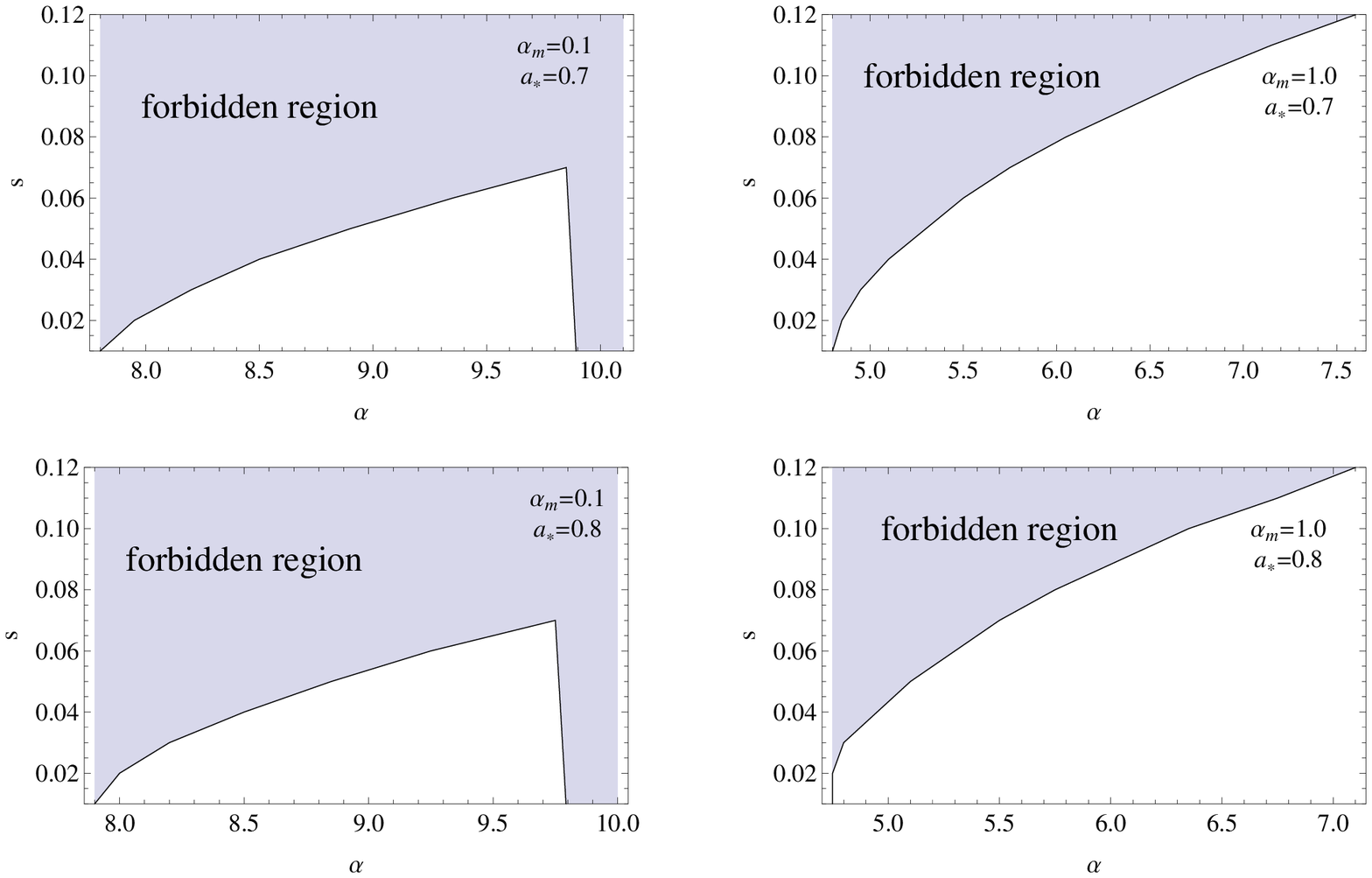}
\caption{The contour of $F_\mathrm{rad}(r_\mathrm{d})=0$ for different values of $a_*$ and $\alpha_\mathrm{m}$ in $s-\alpha$ parameter space, in which $F_\mathrm{rad}(r_\mathrm{d})$ becomes negative in the forbidden region.}
\label{fig5}
\end{figure}

\begin{figure}[ht]
\centering
\includegraphics[width=144mm]{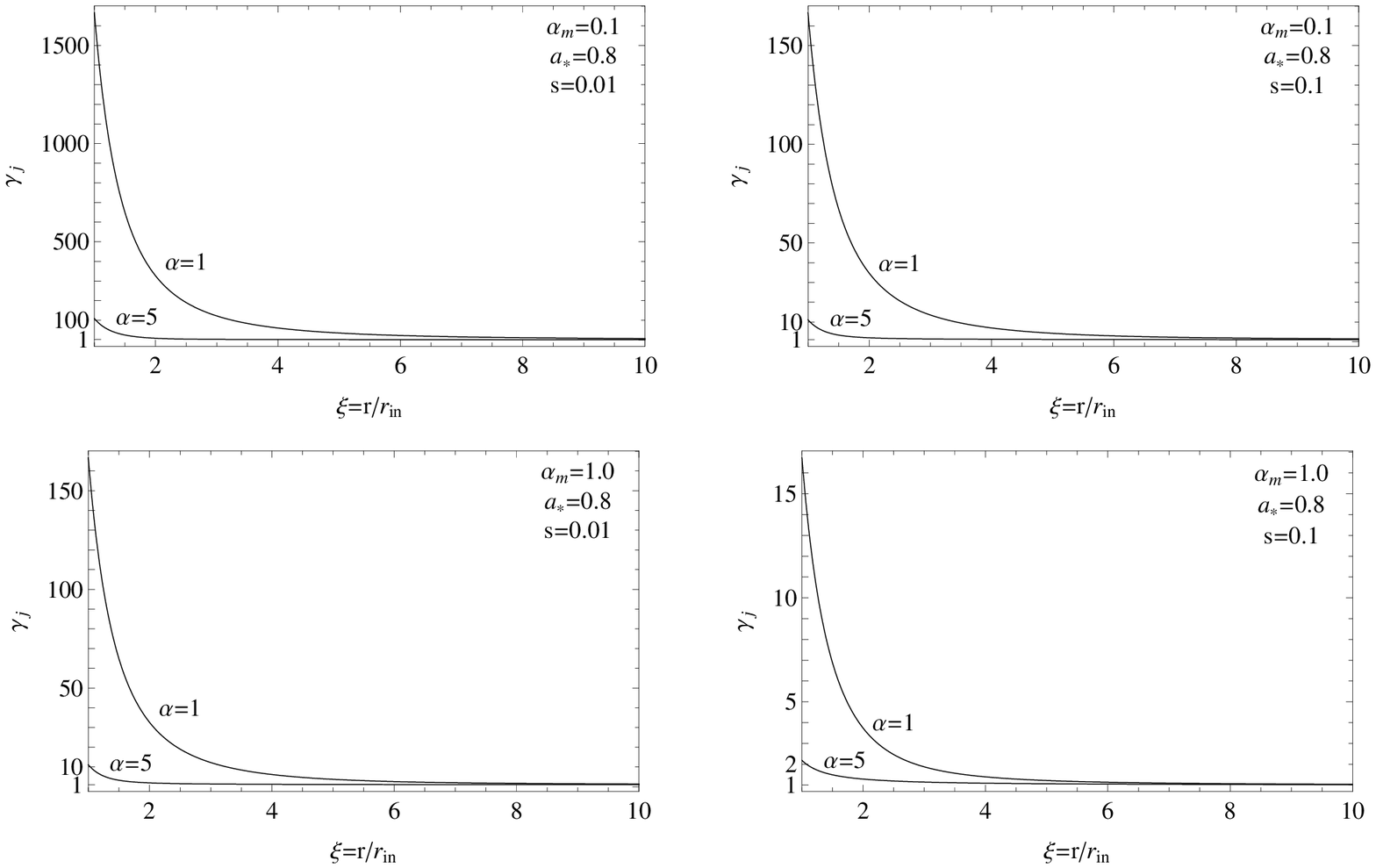}
\caption{The curves of Lorentz factor $\gamma_\mathrm{j}$ versus disk radius $\xi \equiv r_\mathrm{d}/r_\mathrm{in}$ for different values of $s$ and $\alpha_\mathrm{m}$.}
\label{fig6}
\end{figure}

\begin{enumerate}[label=(\roman*), topsep=0pt]
 \item The contour of $F_\mathrm{rad}(r_\mathrm{d})=0$ can be plotted in $\alpha - s$  parameter space by using equation (\ref{eq37})
     as shown in Figure \ref{fig5}, in which $F_\mathrm{rad}(r_\mathrm{d})$ becomes negative in the forbidden region.
 \item The Lorentz factor in the BP process, $\gamma_\mathrm{j}$, can be calculated in our model (see equation(\ref{eq25}) and Appendix for details), and the curves of $\gamma_\mathrm{j}$ varying with disk radius for different values of $\alpha$, $\alpha_\mathrm{m}$ and $s$ are shown in Figure \ref{fig6}. On the other hand, the Lorentz factor $\Gamma_\mathrm{j}$ in LH state should be no greater than 2 (\citealt[hereafter FBG04]{fbg04}). Considering that jet is driven by the BZ and BP processes in our model, and the Lorentz factor of BZ jet is generally greater than that of the BP jet, we have $\gamma_\mathrm{j} < \Gamma_\mathrm{j} \leq 2$. From Figure 6 we conclude that the parameter $\alpha$ should be no less than 5, i.e., $\alpha \geq 5$.
\end{enumerate}

Inspecting Figure \ref{fig5}, we have the constraint of positive disk radiation on the parameters, $\alpha$, $\alpha_\mathrm{m}$, and $s$,
i.e., $4.5 < \alpha < 7$, $\alpha_\mathrm{m}=1$, and a small $s$, such as $s \approx 0.01 \sim 0.02$.

Inspecting Figure \ref{fig6}, we have the constraint of the Lorentz factor on the parameters $\alpha$, $\alpha_\mathrm{m}$, and $s$,
i.e., $\alpha \geq 5$, $\alpha_\mathrm{m}=1$ and $0.01< s <0.1$.

Combining the above results, we can select the values of these parameters in the set ($\alpha_\mathrm{m}=1, \alpha=5, s=0.02$) or
($\alpha_\mathrm{m}=1, \alpha=5, s=0.01$) in fitting the LH states with a steady jet of the four BHXBs as shown in Table \ref{tb1}.

\subsection{Fitting Spectral Profiles of LH State of BHXBs}
\label{sect4.2}

The spectra of the LH state are fitted based on disk-corona model given by \citet[hereafter \citetalias{gwl09}]{gwl09}. This model is different from \citetalias{gwl09} in three aspects. (i) The magnetic field configuration consists of large-scale open field lines threading BH horizon and accretion disk as shown in Figure \ref{fig2}, while that in \citetalias{gwl09} consists of large-scale closed field lines connecting the BH horizon and the inner disk. (ii) The BZ and BP mechanisms are invoked respectively to drive jets from a spinning BH and its surrounding accretion disk, and energy is extracted respectively from the BH and the inner disk to remote astrophysical loads. While in \citetalias{gwl09}, we have no open magnetic field for jet launching, and energy is transferred from the BH into the inner disk. (iii) As in \citetalias{gwl09}, inverse Compton scattering is taken as the radiation process, and Monte Carlo method is used in fitting the spectra of LH state. However, the code used in \citetalias{gwl09} is modified in this case by considering energy transfer into the jet as shown in equation (\ref{eq37}), and the outer boundary of corona is fixed at $40M$ rather than at the outer boundary of the closed field lines in \citetalias{gwl09}.

\begin{figure}[ht]
\centering
\includegraphics[width=120mm]{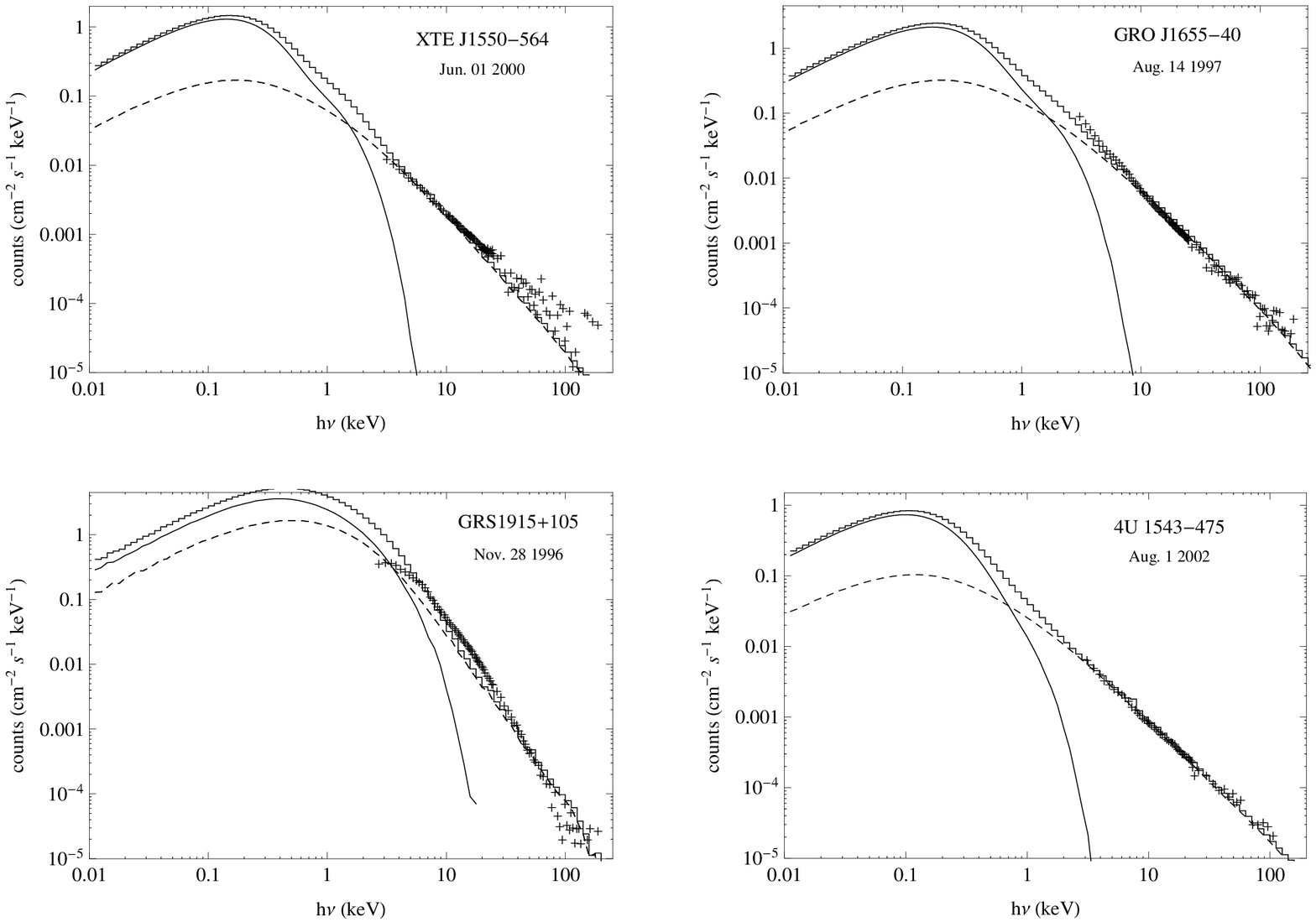}
\caption{The spectral profiles of LH state of four BHXBs are plotted in zigzag lines, which are superposition of
    thermal and power law components in solid and dashed lines, respectively.}
\label{fig7}
\end{figure}

The fitting is carried out based on the features of the four BHXBs taken from \citet[hereafter \citetalias{nm12}]{nm12} as input parameters as shown in Table \ref{tb1}, and the spectral profiles of the LH state are shown in Figure \ref{fig7}.

\begin{table}[ht]
\begin{center}
\caption{Input and fitting parameters of LH state of four BHXBs \label{tb1}}
\begin{tabular}{crrrrrrrr}
\hline\noalign{\smallskip}
    BHXBs & \multicolumn{4}{c}{Input parameters} & \multicolumn{4}{c}{Fitting parameters}\\
    & $a_*$ & $M(M_\odot)$ & $D(kpc)$ & $i(^{\circ})$ & $\dot{m}_\mathrm{in}$ & $\alpha$
    & $\alpha_\mathrm{m}$ & s\\
\hline\noalign{\smallskip}
    XTE J1550-564 & 0.34   & 9.10 & 4.38 & 74.7 & 0.032 & 5 & 1.0 & 0.01\\
    GRO J1655-40  & 0.7     & 6.30 & 3.2  & 70.2 & 0.035 & 5 & 1.0 & 0.02\\
    GRS 1915+105 & 0.975 & 14.0 & 11.0 & 66.0 & 0.200 & 5 & 1.0 & 0.02\\
    4U 1543-475    & 0.8      & 9.4  & 7.5  & 20.7 & 0.005 & 5 & 1.0 & 0.02\\
\noalign{\smallskip}\hline
\end{tabular}
\end{center}
\end{table}

It is noticed that the spectral profiles of the LH states of the four BHXBs given in Figure \ref{fig7}
are in good agreement with the observation data given in Fig. 4.11 of \citetalias{mr06}.

\subsection{A Constraint to Magnetic Field Configuration Based on the Relation between Jet Power
and X-ray luminosity}
\label{sect4.3}

The relation between jet power and X-ray luminosity (hereafter RJPXL) in BHXBs was first
proposed by \citet{fgj03}, and it reads

\begin{equation}
\label{eq38}
L_\mathrm{J}=A_\mathrm{steady}L_\mathrm{X}^{0.5}.
\end{equation}
where the coefficient $A_\mathrm{steady}$ varies between $6\times 10^{-3}$ and $0.3$  (\citetalias{fbg04}; \citealt{mmf04}).

As is well known, the evolution of LH state in one outburst of BHXBs can be depicted in the
X-ray hardness-intensity diagram (HID) as given by \citetalias{fbg04}, and RJPLX implies that the jet
power correlates with the X-ray luminosity in a non-linear way. Since this relation is reduced from
observations, we can regarded it as a constraint on the magnetic field configuration of our model.

In our model $L_\mathrm{J}$ is regarded as $P_\mathrm{jet}$ given by equation (\ref{eq27}), and the values of the
concerned parameters are listed in Table \ref{tb2}, in which the leftmost values of $L_\mathrm{X}$ are calculated
based on the spectral profiles of the LH state given in Figure \ref{fig7}.

\begin{table}[ht]
\begin{center}
\caption{Checking the relation between jet power and X-ray luminosity in LH state \label{tb2}}
\begin{tabular}{ccrrrrr}
\hline\noalign{\smallskip}
    BHXBs & \multicolumn{6}{c}{parameters}\\
\hline\noalign{\smallskip}
    GRO J1655-40  & $\dot{m}_\mathrm{in}$        & 0.035    & 0.04     & 0.045    & 0.05     & 0.055\\
                  & $r_\mathrm{out}$ & 1000     & 10.22    & 7.49     & 6.36     & 5.67\\
                  & $L_\mathrm{X}$   & 0.01359  & 0.01604  & 0.01884  & 0.02176  & 0.02468\\
                  & $L_\mathrm{J}$   & 0.02238  & 0.02431  & 0.02636  & 0.02833  & 0.03016\\
                  & $P_\mathrm{BZ}/P_\mathrm{BP}$ & 0.97 & 1.07 & 1.16 & 1.25 & 1.34\\
    4U 1543-475   & $\dot{m}_\mathrm{in}$        & 0.005    & 0.0055   & 0.006    & 0.0065   & 0.007\\
                  & $r_\mathrm{out}$ & 1000     & 11.31    & 8.00     & 6.63     & 5.82\\
                  & $L_\mathrm{X}$   & 0.004908 & 0.005622 & 0.006423 & 0.007239 & 0.008094\\
                  & $L_\mathrm{J}$   & 0.004187 & 0.004481 & 0.004786 & 0.005087 & 0.005378\\
                  & $P_\mathrm{BZ}/P_\mathrm{BP}$ & 1.46 & 1.57 & 1.66 & 1.74 & 1.83\\
\noalign{\smallskip}\hline
\end{tabular}
\end{center}
\end{table}

In Table 2, the radius $r_\mathrm{out}$ represents the outer boundary of the BP magnetic field configuration, and the luminosities and accretion rates are defined in terms of Eddington luminosity and Eddington accretion rate, respectively. As shown in Table \ref{tb2}, the radius $r_\mathrm{out}$ of the outer boundary of the BP magnetic field configuration decreases monotonously with the increasing accretion rate $\dot{m}_\mathrm{in}$, jet power $L_\mathrm{J}$ and X-ray luminosity $L_\mathrm{X}$. This result implies that the magnetic field configuration could be related to the state transitions of BHXBs, and this issue will be discussed in the next section.

\section{DISCUSSION}
\label{sect5}

In this paper, we propose a corona-disk model for fitting the LH state associated with steady
jet of BHXBs based on the magnetic field configuration of the coexistence of the BZ and BP
processes, and some issues related to our model are discussed in this section.

\subsection{Transition from LH to VH States in BHXBs}
\label{sect5.1}

Up to now a consensus on the classification of spectral states of BHXBs has not been reached.
It is widely accepted that the spectral states of BHXBs can be reduced to two basic states, i.e., a
hard state and a soft state \citepalias{mr06}. As shown in HID, X-ray luminosity always increases after an
outburst starts, attaining its maximum in ‘intermediate’ state in the transition from hard to soft
states. However, the properties of the ‘intermediate’ state remain unclear, and different definitions
have been presented, e.g., Steep Power Law (SPL) state by \citetalias{mr06}, ’very high’ (VH) state by \citet{emn97}. \citet{b06} classified ‘intermediate’ state as hard intermediate (HIM) and soft intermediate (SIM) states. In this paper, we take the ‘intermediate’ state as VH
state as given in \citetalias{nm12}, which is associated with the episodic, relativistic jet.

As is well known, state transition in BHXBs display a variety of variations not only in
luminosities but also in some spectral characteristics such as hardness and spectral index. The
complexity is particularly attractive in the transition from hard to soft states, with which different
remarkable phenomena are associated. A visualized description for the main features of state
transitions of BHXBs is given in HID, where the typical spectral evolution traces along a q-shaped
pattern and forms an anti-clockwise cycle (\citealt{b04}; \citealt{bmm11}; \citealt{fb12}; \citetalias{fbg04};
\citealt{fhb09}; \citealt{hb05}). Based on HID the outbursts of
BHXBs are generally triggered by a sudden increase of accretion rate from ‘quiescence’ to LH
state, and the spectra are normally hard with photon index $\sim 1.7$, being associated with steady jets
in LH states, and the jet power is correlated with the X-ray luminosity as $L_\mathrm{J}\propto L_\mathrm{X}^{0.5}$. After reaching
the peak luminosity, the spectra begin to soften and the jets transit from steady into episodic,
indicating the transition from LH state to VH state. After crossing the jet line in HID the VH state
transits to HS state, calming down with soft spectra without jets. The latest research shows that
the HS state associates with a strong disk wind. Finally, BHXB returns to its quiescent state with a
hard spectrum accompanied with the reappearance of jets (\citealt{fb12}; \citealt{z13}).

The variation of the X-ray luminosity and spectra is interpreted naturally by the
corresponding variation of accretion rate and accretion geometry (\citealt{emn97}; \citealt{d02,d10}; \citealt{dgk07}).
A series of works on the formation and evolution of the corona give a physical explanation of the spectral state transitions (\citealt{lmm05}; \citealt{mlm05, mlm09}, see \citealt{z13} for a review).

However, accretion rate is not the only parameter for governing the state transition of BHXBs,
and some phenomena involved cannot be interpreted by only changing accretion rate. For example,
state transition from hard to soft occurs at luminosity higher than that in later reverse transition in
one outburst, this hysteresis cannot be interpreted by the variation of accretion rate (\citealt{mkhe95}; \citetalias{fbg04}; \citealt{b10}).

It was suggested by \citet{su05} that the size of the central magnetic flux
bundle can be identified with the second parameter for determining X-ray spectral states of
BHXBs and the presence of relativistic outflows. Very recently, \citet{k12} pointed out that
the magnetic field might be primarily toroidal in the soft state, but primarily poloidal in the hard
state. In fact, both the accumulation of the magnetic flux in the inner disk and the change between
toroidal and poloidal magnetic fields can be regarded as evolution of magnetic field configuration.
Thus we suggest that magnetic field configuration on the accretion disk could be regarded as the
second parameter for governing the state transition of BHXBs.

\begin{figure}[ht]
\centering
\includegraphics[width=120mm]{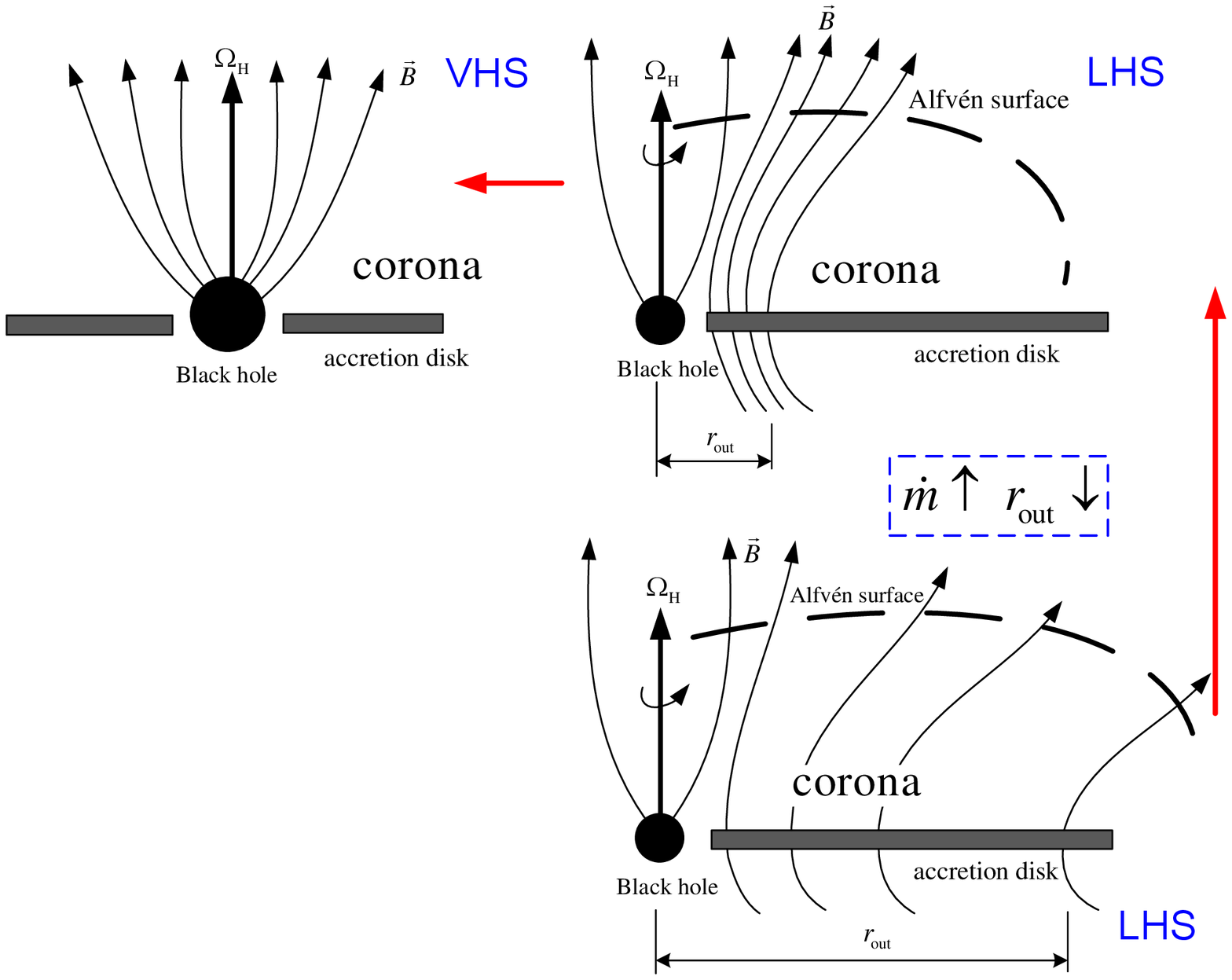}
\caption{A schematic drawing of magnetic field configurations in transition from LH state to VH state in
    BHXBs.}
\label{fig8}
\end{figure}

This viewpoint is strengthened by the constraint of RJPXL on the outer boundary of the BP
magnetic field configuration as shown in Table \ref{tb2}. The correlation of magnetic field
configurations with the transition from LH to VH states is illustrated from bottom-right to
top-left panels in Figure \ref{fig8}, in which the outer boundary of the BP magnetic field configuration
decreases monotonically with the increasing accretion rate $\dot{m}_\mathrm{in}$, $L_\mathrm{J}$ and $L_\mathrm{X}$ for the validity of
RJPXL in LH states of BHXBs given by equation (\ref{eq38}), and the VH state appears as all large-scale
poloidal magnetic fields are carried onto the BH as shown by the top-left panel in Figure \ref{fig8}.

The scenario of evolution of magnetic field configuration is also helpful to understand the
correlation of jet power with BH spin, which has been addressed by a number of authors
(\citealt{m99, mg04, hea04, dhkh05, hk06, lwg08, wcw11}).

Recently, \citet[hereafter \citetalias{fgr10}]{fgr10} pointed out that no evidence for any correlation between the jet powers and the BH spin based on the reported measurements of BH spin and jet power for BHXBs. On the contrary, it was showed in \citetalias{nm12} that the 5-GHz radio flux
of transient ballistic jets in BHXBs correlates with the BH spin estimated via the continuum-fitting method, and they pointed out that it is the first direct evidence of jets powered by BH spin energy.

According to our model the BZ power is not dominant over the BP power in LH state
corresponding to the magnetic field configuration with great outer boundary radius $r_\mathrm{out}$, and it
becomes gradually dominant over the BP power in the transition from LH to VH states with the
decreasing $r_\mathrm{out}$ as shown in Figure \ref{fig8}. It is the magnetic field concentrated on the BH horizon that
results in the jet power proportional to the square of BH spin in VH state. In addition, the transient
ballistic jet in VH state can be interpreted by invoking the kink instability related to the BZ
process (\citealt{wly06}). Therefore by invoking the variation of the large-scale magnetic field
configuration, we can resolve the debate between \citetalias{fgr10} and \citetalias{nm12} on the issue of the jet power
and the BH spin in BHXBs.

\subsection{Energy Conversion in Jet Launcing and Corona Current}
\label{sect5.2}

\begin{figure}[ht]
\centering
\includegraphics[width=120mm]{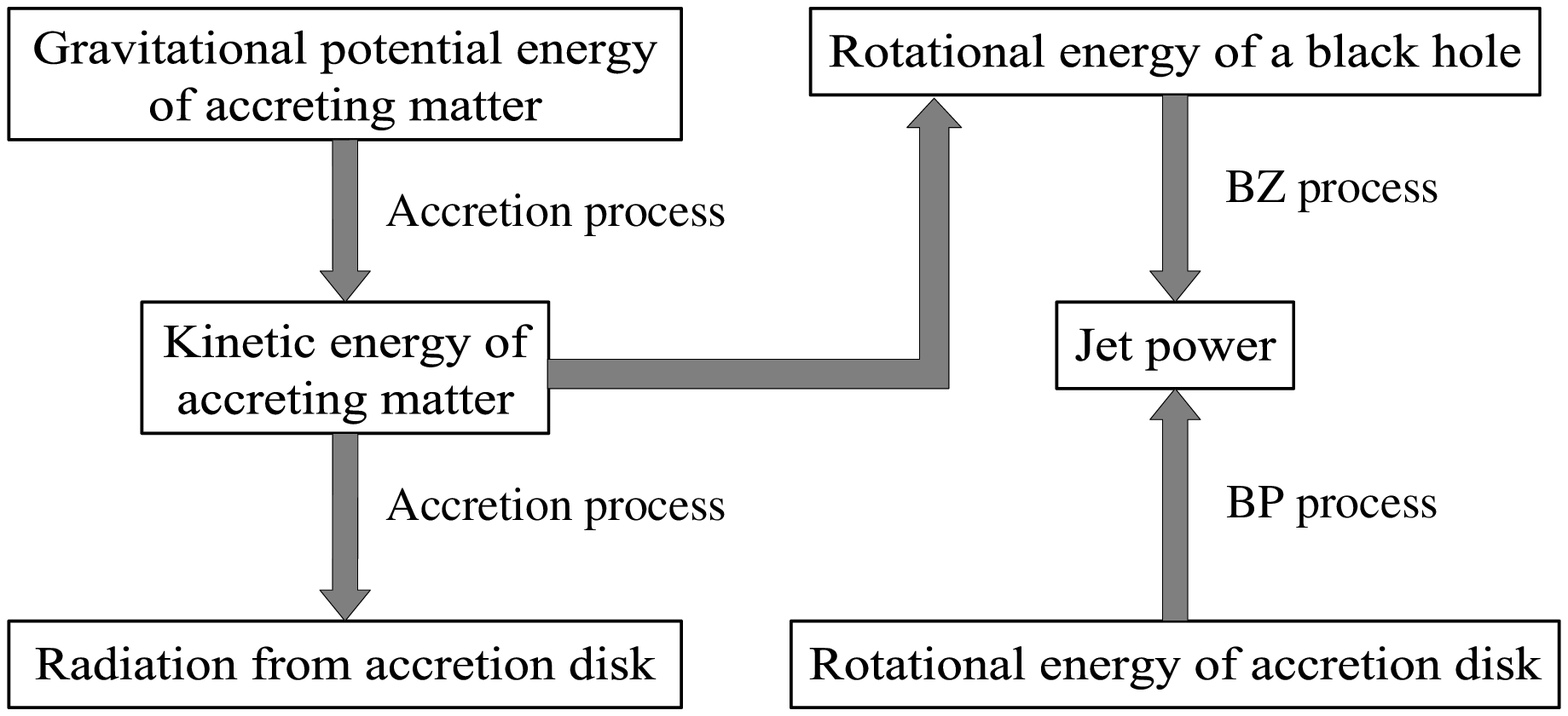}
\caption{A block diagram of energy release and conversion in accretion disk with BZ and BP processes.}
\label{fig9}
\end{figure}

In our model, energy is released from two sources: (i) rotational energy from a spinning BH
via the BZ process and (ii) rotational energy from disk via accretion process with the BP process.
Energy release and conversion are illustrated in Figure \ref{fig9}.

Energy release and conversion are outlined in Figure \ref{fig9}. It is shown that two energy sources
(gravitational potential energy of accreting matter and rotational energy of a black hole) give rise
to two types of energy output from BH system, i.e., radiation via accretion process and jet power
via BZ and BP processes are included. Obviously, both magnetic field and rotational energy of a
BH arise from accretion process, so accretion process is essential for the BZ process.

As to energy conversion in the jet, we introduce corona current, which is required by
continuity of current flowing on the disk as shown in Figure \ref{fig3}. Similarly, corona current is also
essential for energy conversion in the BZ jet, being required by continuity of current flowing on
the stretched horizon of a spinning black hole (\citealt{tpm86}).

In addition, corona current could be related to the following issues. (i) Strengthening the
toroidal magnetic field, being essential for Poynting flux near the disk surface as shown in Figure
\ref{fig3}; (ii) an alternative way of enhancing corona temperature in the form of Joule heating; (iii) an
alternative way of exchanging energy between disk and corona. We shall discuss these issues in
our future work.

\subsection{Advantages and Disadvantage of This Model}
\label{sect5.3}

Compared to the widely believed model (ADAF) the advantages of our model are related to
jet launching and its application to fitting LH states of BHXBs, which are summarized as follows.

\begin{enumerate}[label=(\roman*), topsep=0pt]
 \item Required by the energy conversion from Poynting flux to the kinetic energy flux in the jet from accretion disk, coronal current flowing across the magnetic surfaces is introduced naturally in this model. Corona current is essential for continuity of current flowing on the accretion disk, being crucial for launching a steady jet via the BP process.
 \item Based on energy conversion in the jet and the work done by magnetic torque exerted on disk current and corona current, we construct a large-scale magnetic field configuration on the disk for jet launching, and the LH state is fitted by invoking accretion process with the coexistence of the BZ and BP processes.
 \item Based on above magnetic field configuration, we discuss the relative importance of BZ to BP powers in terms of a few parameters constrained by observational and theoretical considerations, and apply this result to fit the LH state associated with a steady jet.
 \item Required by the validity of RJPXL, we find that the outer boundary of the BP magnetic field decreases monotonously with the increasing jet power and X-ray luminosity in LH states, and this implies that magnetic field configuration could be regarded as the second parameter for governing the transition from hard to soft states in BHXBs.
\end{enumerate}

On the other hand, there exist some disadvantages with this model, being given as follows.

\begin{enumerate}[label=(\roman*), topsep=0pt]
 \item Although corona current is introduced based on some reasonable consideration, we have not presented a detailed analysis on it, such as how corona current distributes in the corona, and how it interacts with the disk, and how it affects the radiation or spectrum, etc.
 \item Only inverse Compton scattering is taken into account as radiation mechanism in fitting the spectra of LH states as a simplified model. As a matter of fact, synchrotron radiation or SSC might be important in fitting. Likewise, we didn’t consider the contribution of jet to the radiation.
 \item We fail to discuss hysteresis in state transition of some BHXBs, which involves a higher luminosity at the transition from hard to soft spectral states and a lower one at the reverse transition from soft to hard spectral state. Although explanation has been given by disk evaporation model (e.g. \citealt{mlm05}), the physics behind hysteresis remains elusive.
\end{enumerate}

We hope to overcome the above disadvantages and modify this model in future work.

\normalem
\begin{acknowledgements}

We are very grateful to the anonymous referee for his (her) helpful comments on the manuscript.
This work is supported by the National Basic Research Program of China (2009CB824800) and
the National Natural Science Foundation of China under Grant No. 11173011.

\end{acknowledgements}

\appendix                  

\section{APPENDIX: DERIVATION OF EQUATION (25)}

\citetalias{c02} gives the mass loss rate in the jet from unit surface area of a disk as follows,
\begin{equation}
\label{a1}
\dot{m}_\mathrm{jet}=\frac{(B_\mathrm{d}^\mathrm{P})^2}{4\pi}(r_\mathrm{d}\Omega_\mathrm{d})^\alpha
\frac{\gamma_\mathrm{j}^\alpha}{(\gamma_\mathrm{j}^2-1)^{\frac{\alpha+1}{2}}}.
\end{equation}

According to equation (\ref{eq17}) and the context, we have
\begin{equation}
\label{a2}
\dot{M}_\mathrm{jet}=4\pi r_\mathrm{d}\dot{m}_\mathrm{jet}=r_\mathrm{d}(B_\mathrm{d}^\mathrm{P})^2
(r_\mathrm{d}\Omega_\mathrm{d})^\alpha\frac{\gamma_\mathrm{j}^\alpha}{(\gamma_\mathrm{j}^2-1)^{\frac{\alpha+1}{2}}},
\end{equation}
Combining equations (\ref{eq17}) and (\ref{eq22}), we have
\begin{equation}
\label{a3}
B_\mathrm{d}^\mathrm{P}=B_\mathrm{in}\left(\frac{r_\mathrm{d}}{r_\mathrm{in}}\right)^{s-2}
=\sqrt{\frac{\dot{M}_\mathrm{in}}{\alpha_\mathrm{m}r_\mathrm{H}^2}}\left(\frac{r_\mathrm{d}}{r_\mathrm{in}}\right)^{s-2},
\end{equation}
Incorporating equations (\ref{a2}), (\ref{a3}) and (\ref{eq17}), we have
\begin{equation}
\label{a4}
\frac{1}{\alpha_\mathrm{m}r_\mathrm{H}^2}\left(\frac{r_\mathrm{d}}{r_\mathrm{in}}\right)^{s-2}\frac{r_\mathrm{in}^2}{s}
(r_\mathrm{d}\Omega_\mathrm{d})^\alpha=\frac{(\gamma_\mathrm{j}^2-1)^{\frac{\alpha+1}{2}}}{\gamma_\mathrm{j}^\alpha},
\end{equation}
And equation (\ref{eq25}) is the dimensionless form of equation (\ref{a4}).

\end{document}